\documentclass[sigconf]{acmart}

\renewcommand\footnotetextcopyrightpermission[1]{} % removes footnote with conference information in first column
\usepackage{booktabs} % For formal tables
\usepackage{multirow}
\usepackage{algorithm}
\usepackage{algpseudocode}
\usepackage{calrsfs}
\usepackage{balance}
\usepackage[symbol]{footmisc}

\makeatletter
\def\@copyrightspace{\relax}
\makeatother

\begin{document}

\fancyhead{}

\copyrightyear{2020} 
\acmYear{2020} 
%\setcopyright{draft}
\setcopyright{acmlicensed}
\acmConference[Draft]{}{June 15, 2020}{}
%\acmConference[AdKDD]{AdKDD, The 25th ACM SIGKDD International Conference on Knowledge Discovery \& Data Mining}{August 23, 2020}{San Diego, California, USA}
%\acmPrice{15.00}
%\acmDOI{10.1145/1234567.89012345}
%\acmISBN{978-1-4503-5552-0/18/08}

\title{Digital Contact Tracing Using IP Colocation}

\def\sharedaffiliation{\end{tabular}\newline\begin{tabular}{c}}
\def\wu{\superscript{*}}
\def\wp{\superscript{+}}
\def\wg{\superscript{\dag}}

% AUTHORS
% ---------------------

\author{Matthew Malloy}
\affiliation{
	\institution{University of Wisconsin}
}
\email{matthew.malloy@wisc.edu}

\author{Aaron Cahn}
\affiliation{
	\institution{Comscore}
}
\email{acahn@comscore.com}

\author{Jon Koller}
\affiliation{
	\institution{Comscore}
}
\email{jkoller@comscore.com}

\renewcommand{\shortauthors}{}

\begin{abstract}
%Internet devices such as PCs, smartphones, and tablets present unique identifiers (\emph{e.g.}, web cookies or advertising IDs) when they access resources on the internet.
%These IDs are used in a variety of ways, including targeted advertising, tracking, measurement, and reporting.
%Browsing across devices, and the variety of applications that are used to access the internet, results in a multiplicity of IDs associated with single devices and users.

The spread of an infectious disease through a population can be modeled using a network or a graph.  In digital advertising, \emph{internet device graphs} are graph data sets that organize identifiers produced by mobile phones, PCs, TVs, and tablets as they access media on the internet.  Characterized by immense scale, they have become ubiquitous as they enable targeted advertising, content customization and tracking.  This paper posits that internet device graphs, in particular those based on \emph{IP colocation}, can provide significant utility in predicting and modeling the spread of infectious disease.  Starting the week of March 16th, 2020, in the United States, many individuals began to `shelter-in-place' as schools and workplaces across the nation closed because of the COVID-19 pandemic.  This paper quantifies the effect of the shelter-in-place orders on a large scale internet device graph with more than a billion nodes by studying the graph before and after orders went into effect.  The effects are clearly visible.   The structure of the graph suggests behavior least conducive to transmission of infection occurred in the US between April 12th and 19th, 2020.  This paper also discusses the utility of device graphs for \emph{i)} contact tracing, \emph{ii)} prediction of `hot spots',  \emph{iii)} simulation of infectious disease spread, and \emph{iv)} delivery of advertisement-based warnings to potentially exposed individuals.  The paper also posits an overarching question: can systems and datasets amassed by entities in the digital ad ecosystem aid in the fight against COVID-19?

%A popular mechanism for generation of device graphs is \emnph{IP-colocation}, which relies on the observation of two users as they access the internet from the same WiFi access point.  As the shelter in place orders went into effect, users often connected to fewer WiFi access points. 

\end{abstract}


%\ccsdesc[300]{General and reference~Measurement}
%\ccsdesc[300]{Mathematics of computing~Bayesian computation}
%\ccsdesc[300]{Information systems~Online advertising}
%\ccsdesc[300]{Computing methodologies~Supervised learning by classification}


\maketitle

\section{Introduction} \label{sec:introduction} Epidemiologists often study the spread of infectious disease using a \emph{contact process} on a network (i.e, a graph) \cite{danon2011networks, khim2018theory, anderson1992infectious, christakis2007spread, dong2012graph}.  In a contact process, an infection spreads along the edges of the graph, infecting neighboring nodes.  While the individual edges define who becomes infected, the underlying structure of the graph controls the characteristics of the spread of the infectious disease through the population.   Knowledge of the edges in the network is paramount to contact tracing. 
 
In digital advertising, \emph{device graphs} are datasets that relate digital identifiers produced by smartphones, PCs and tablets as they access resources on the internet.  Device graphs, characterized by immense scale with billions of nodes and tens of billions of edges \cite{Malloy19}, have become ubiquitous in the digital advertising ecosystem as they facilitate targeted advertising.  While device graphs are constructed using information gathered online, the associations they capture correspond to offline relationships. %, including social and contact relationships. 

While the exact mechanisms for assembling device graphs vary, the principles of their construction follow a similar paradigm.   Websites, apps and advertising networks log information about smarphones and PCs as they access media on the internet.  The information they collect includes a unique, anonymized identifier (either a \emph{cookie}-ID or an advertising ID), an IP address, a timestamp, and information about the device and the content accessed. 

One technique for generating device graphs is \emph{IP colocation}.  IP colocation is the co-occurrence of identifiers on a single IP address at approximately the same time.  The efficacy of IP colocation techniques comes from the observation that IP space is \emph{intimate}:  IP addresses are shared by family members, friends and co-workers when they connect to the internet via the same WiFi access point.  Generally speaking, devices that connect to a WiFi router have the same public-facing IPv4 address.  Capturing these relationships is valuable for marketing and targeted advertising, and many entities in digital advertising have assembled device graphs \cite{Tapad, Drawbridge, Lotame}.  These datasets cover the majority of internet-connected devices in the United States and elsewhere in the world \cite{Malloy17}.  

While internet device graphs are ubiquitous in digital advertising, they also have potential utility for epidemiological study and combating the worldwide COVID-19 pandemic.  As WiFi access points have limited range, IP colocation is a proxy for \emph{physical colocation}.  As physical colocation supposes \emph{epidemiological contact}, device graphs are a natural fit to study the spread of infectious disease over a network.  This paper posits that device graphs could be used to:  \emph{i)} supplement \emph{contact tracing} by enabling public health officials to identify users that connected to the same WiFi access points at internet scale, \emph{ii)} enable prediction of outbreak `hot spots',  and \emph{iii)} facilitate delivery of advertisement-based warnings to potentially exposed individuals.  While the datasets have great potential, there are also limitations.  Unlike digital contact tracing approaches based on mobile apps and technology such as GPS and Bluetooth Low Energy \cite{ferretti2020quantifying, gorji2020stecc}, IP colocation is more susceptible to false positives.  This potential shortcoming is arguably overcome by scale. 

This paper studies the effects of \emph{shelter-in-place} on an internet scale IP colocation device graph provided by a large internet analytics company, Comscore.  The COVID-19 pandemic caused a significant change in day-to-day behavior in the US and across the globe.  In the US, starting approximately the week of Monday March 16th, many individuals began to work from home and shelter-in-place either voluntarily or through mandate, as schools across the nation closed.  We study the effects of this change in behavior by examining the community structure of the device graph\footnote{We intend to  make anonymized graph datasets publicly available, corresponding to before and after the shelter-in-place orders. }.  

The effects of sheltering in place are clearly visible in the device graph, validating a simple hypothesis: when individuals stay home, they do not connect to WiFi outside-the-home, reducing the number of outside the home edges in the graph.  We study two statistics of the graph and the inferred community structure: \emph{i)} the modularity, and \emph{ii)} the ratio of edges that cross communities to the total number of edges in the graph.  Both metrics indicate that behavior least conducive to the spread of infectious disease occurred on the week beginning April 12th, 2020.   Past this week, both graph metrics suggest behavior becomes more conducive to the spread of infectious disease.  We also compare the metrics to the estimated percent of the US population that is placed under stay at home orders.

The use of internet device graphs for contact tracing, prediction of transmission, and epidemiological studies, to the best of our knowledge, is a new idea.  Compared to other technology for digital contact tracing using mobile devices \cite{ferretti2020quantifying}, there are both significant advantages and limitations.  One main advantage of device graphs is that they already exist and do not require custom apps or websites as are often suggested for digital contact tracing in the media \cite{nytimes_contact_tracing}.  Required data collection only consists of server logs containing an IP address, an ID (a cookie or advertising ID), and a timestamp.  This data is recorded any time a user accesses content on a webpage or app, and these datasets are already produced and sold by a number of tracking and measurement entities in the digital advertising ecosystem.  The identifiers used in device graphing conform to established privacy guidelines, and are also precisely the identifiers that facilitate targeted advertisements, which could be used for public service announcements or warnings of potential exposure to infectious diseases.   There are limitations to using device graphs based on IP colocation for contact tracing.  Most importantly, they do not reveal detailed location traces that are available using techniques based on precise geolocation services (i.e, GPS).  Instead they provide reliable estimation of habitual behavior.

To the best of our knowledge, the use of device graphs to simulate the spread of infectious disease is a new idea.  The baseline graph under normal conditions would likely be of interest to epidemiologists for use in a contact process model.  Moreover, the change in characteristics of the graph pre and post shelter-in-place orders is also likely of interest to epidemiologists and policy makers in understanding the effectiveness of such orders.  The data structure provides a unique opportunity for simulation and exposition to answer the following question: do the shelter-in-place orders have material impact on the rate of spread of infectious disease as modeled as a contact process on a device graph? A number of related modeling questions ensue.  We also pose the overarching question: can the processes and datasets developed by the advertising ecosystem writ large aid in the fight against COVID-19? 
%In particular, will the effect of shelter in place impact the reproduction number $R_0$ as simulated on the graph?

\section{Background and Methodology} \label{sec:methodology} This section reviews background and methodology for generation of an IP colocation device.  More details can be found in \cite{Malloy17, Malloy18, Malloy19}.  IP colocation device graphing often follows three steps: \emph{i)} data collection, \emph{ii)} generation of the graph, which establishes \emph{pair-wise} relationships between devices by observing them on the same IP address at approximately the same time, and \emph{iii)} community detection, which partitions the graph into smaller clusters that identify \emph{household, person-level, or larger internet communities} depending on tuning of the algorithms.  

Before proceeding, we define notation required in graph analysis.  An undirected, weighted graph $G$ is a set of nodes $V$ and a set of edges $E$,  $G = (V, E)$.  An edge $e \in E$ consists of two elements from $V$ and a weight: $e = (i, j, w) \in V \times V \times \mathbb{R}$.  In a \emph{device} graph, a node $i \in V$ is a digital identifier ({\em e.g.,} a web cookie or advertising ID).   An edge $e \in E$ represents a relationship between two identifiers.  Many of the applications of device graphs require determining groups of devices that have strong relationships in the graph.  We refer to these groups as {\em communities}.  When a community is aligned to a residential household, we refer to it as a household.   A \emph{community} is a set of one or more nodes, $C = \left\{i, \dots \right\} \subset V$, and the set of communities is denoted $\mathcal{C} = \{ C_1, C_2 \dots \}$.  $\mathcal{C}$ is a partitioning of $V$, i.e, $C_i \cap C_j = \{\}$ for any $i \neq j$, and $\bigcup_i  C_i = V$.  In IP colocation device graphs, communities are small groups of IDs often observed together on the same IP address. These groups are are well aligned with residential households and small business places.

%\emph{IP-colocation} is the co-occurrence of requests from two device IDs from the same IP address in a relatively short time span, and is the basis for the graph presented in this paper.    The resulting graph is deterministic in identifying relationships based on IP address colocation but probabilistic in identifying household clusters as described below.

The data used to build the IP colocation device graphs comes from Comscore's digital network, one of the largest in the world. This data is collected over a 42 day period (six weeks) via the local execution of either a JavaScript/HTML or SDK (software development kit) tag on a client machine. These tags can be found accompanying a wide variety of internet resources, including web pages, video requests, mobile applications, advertisement deliveries, and other distributed content. When local execution occurs a unique record is directly reported to Comscore's infrastructure. The record is run through an ETL (extract-transform-load) process before being utilized in the construction of the IP colocation device graph.

Construction of the IP colocation device graph starts with a dataset consisting of tuples of \emph{(device ID, IP address, time)}.  For each IP address in the dataset, an edge is established between every pair of device IDs that share that IP address on epoch $t=1$.   The weight of the edge is inversely proportional to the number of digital identifiers observed on that IP at that epoch.  This is repeated for epoch $t=2,3,\dots$.  After each epoch is considered, a final weight is assigned to each pair of devices by summing over all epochs and all IP addresses.  The algorithm is detailed in Algorithm \ref{alg:graph}. Algorithm \ref{alg:graph} requires two parameters: $N_{\mathrm{max}}$, which is set to exclude high volume IP addresses, and the \emph{edge cutoff} parameter $\gamma$, which results in the exclusion of edges below a specified weight. Algorithm \ref{alg:graph} is implemented in practice utilizing the Apache Pig platform on an Apache Hadoop environment consisting of 500+ worker nodes.

%\textcolor{red}{Add discussion of processing environment}

%\begin{algorithm}
%\caption{IP Colocation Device Graph}\label{alg:graph}
%\begin{algorithmic}[1]
%\State {\bf{input}} tuples \emph{(device ID, IP address, time)}
%\State {\bf{initialize}} $E = \{ \}$
%\State {\bf{for}} each time $t$, each IP $k$
%\State \hspace{.4cm} $N_{k,t} =$ distinct devices on $k$ at time $t$
%\State \hspace{.4cm} {\bf{if}} $N_{k,t} \leq N_{\mathrm{max}}$
%\State \hspace{.8cm} {\bf{for}} each pair of devices $i,j$ on IP $k$ 
%\State \hspace{1.2cm} $E  = E \cup \{(i,j)\}$
%\State \hspace{1.2cm} $w_{i,j}(k,t) =  \frac{1}{N_{k,t}}$
%\State \textbf{return} $E$,  $w_{i,j} = \sum_{t,k} w_{i,j} (t, k)$ for all $ (i,j) \in E$
%\end{algorithmic}
%\end{algorithm}

%\begin{algorithm}
%\caption{IP Colocation Device Graph}\label{alg:graph}
%\begin{algorithmic}[1]
%\State{{\bf{parameters}}: $N_{\mathrm{max}}$, $\gamma$}
%\State {\bf{input}}: tuples \emph{(device ID, IP address, time)}
%\State {$V= $ unique device s}
%\State {\bf{for}} each time $t$, each IP $k$
%\State \hspace{.4cm} $\pazocal{S}_{t,k}$ := \{set of devices on IP $k$ at time $t$ \} 
%\State \hspace{.4cm} {\bf{if}} $|\pazocal{S}_{t,k}| \leq N_{\mathrm{max}}$
%\State \hspace{.8cm} {\bf{for}} each pair $(i,j)$ on IP $k$ at $t$
%\State \hspace{1.2cm} $w_{i,j}(t,k) =  \frac{1}{ |\pazocal{S}_{t,k}| }$
%\State $w_{i,j} = \sum_{t,k} w_{i,j} (t, k)$ for all $(i,j)$
%\State $E = \{(i,j,w_{i,j} ) :  w_{i,j}  > \gamma \}$ 
%\State \textbf{return} $G = (V,E)$
%\end{algorithmic}
%\end{algorithm}

\begin{algorithm}
\caption{IP Colocation Device Graph   \cite{Malloy17} }\label{alg:graph}
\begin{algorithmic}[1]
\State{{\bf{parameters}}: $N_{\mathrm{max}}$, $\gamma$}
\State {\bf{input}}: tuples \emph{(device ID, IP address, time)}
\State {$V= $ set of unique device IDs}
\State {\bf{for}} each time step $t$, each IP $k$
\State \hspace{.4cm} $N_{t,k} = $ number of distinct device IDs on IP $k$ at time $t$
\State \hspace{.4cm} {\bf{if}} $N_{t,k} \leq N_{\mathrm{max}}$
\State \hspace{.8cm} {\bf{for}} all pairs of device IDs $(i,j)$ on IP $k$ at $t$
\State \hspace{1.2cm} $w_{i,j}(t,k) =  \frac{1}{ N_{t,k} }$
\State $w_{i,j} = \sum_{t,k} w_{i,j} (t, k)$ for all $(i,j)$
\State $E = \{(i,j,w_{i,j} ) :  w_{i,j}  > \gamma \}$ 
\State \textbf{return} $G_{\gamma} = (V,E)$
\end{algorithmic}
\end{algorithm}

After construction of the graph, community detection algorithms can be applied to cluster devices into small communities of closely related nodes.  The clusters, which partition $V$, correspond to individuals, residential households, and larger work and social communities.  As the basis for the graph is IP colocation, the communities correspond to individuals that share WiFi access points.  To study clustering properties of the graph, we applied community detection algorithms, adjusting the parameter $\gamma$ as one mechanism to control the size of the communities.  

The goal of community detection is to find a partitioning of the nodes so that the edges between nodes in a group are dense and edges between nodes from different groups are sparse.  Community detection is well studied with many available methods.  The Louvain method  ~\cite{Blondel08} is a popular technique due to its success on large scale graphs, and we employ the approach to cluster the nodes of the IP colocation graph.   The Louvain approach consists of two repeated steps - \emph{i)} minimizing a cost function by moving nodes to neighboring communities and \emph{ii)} creating a new graph with the communities from the first step as nodes.  The technique creates a hierarchy of communities, each corresponding to a different partitioning of the node set.  In this paper we focus only on communities created after the first iteration, i.e, step \emph{i)} above.

While the resulting communities are based on relationships in IP space, their interpretation translates into offline relations between devices.  The parameter $\gamma$ is a cutoff for the minimum edge weight in the graph.  Tuning $\gamma$ also has an effect on community size.  For small $\gamma$, the communities correspond to multiple identifiers that belong to the same person.  As $\gamma$ is decreased, larger communities are produced that align with groups of devices that share a residential household. Further decreasing $\gamma$ generates \emph{internet communities} that connect extended families, friends and co-workers.  In this paper we focus on parameter settings that result in communities that corresponding to residential households.

\section{Results} \label{sec:results} We study the characteristics of the graph and the communities, focusing on two metrics: graph modularity, and the ratio of edges that \emph{cross} communities to the total number of edges.   Modularity, denoted $Q$, can be expressed as \cite{brandes2007finding}
\begin{eqnarray*}
 Q = \sum_{s} \left(\frac{\ell_s}{L} - \left(\frac{d_s}{2L}\right)^2  \right)
 \end{eqnarray*} 
where $s$ enumerates the communities, $\ell_s$ is the number of edges inside the community, $d_s$ is the total degree of the nodes in community $s$, $N = |V| $, and  $L = |E| $.  The ratio of edges inside communities to those outside is expressed as:
\begin{eqnarray*}
R = \frac{E - \sum_{s} \ell_s  }{E}.
\end{eqnarray*}
Both metrics quantify how well the graph conforms to the community structure.  Graph modularity can be interpreted as the ratio of the number of edges inside a community, minus the expected number of edges if connections were defined at random, to the total number of edges.  Graphs with fully connected communities and no cross community edges have a modularity $Q =1$, and graphs with no community structure and uniformly random edges over the nodes have a modularity of $Q=0$. 

To analyze the effect of stay-at-home orders across the United States on both graph modularity and the ratio of edges that cross communities to the total number of edges, we looked at a series of graphs produced using Comscore data collected over several months. All data examined originated from the US between the dates of UTC January 20, 2020 and UTC June 7, 2020 - a 140 day period (20 weeks).  Each graph consists of 6 weeks of data.   The date associated with the graph is the \emph{center} date of the time frame; e.g., a device graph on April 1st consists of data collected the three weeks immediately preceding and following April 1st. In the hierarchy of partitions produced by Louvain Modularity, only results from the finest scale communities are presented. All results are presented for $G_{0.8}$. 
 
As the United States began to stay-at-home, the frequency with which the population shared IP space with co-workers, friends, and others outside of their home decreased.  The main hypothesis is that such orders lead to a decrease in cross community edges and an increase in graph modularity.  Figures \ref{fig:modularity} and \ref{fig:intra} confirm this hypothesis.  Both metrics indicate the graph is most modular on the week beginning April 12th. This suggests the US in aggregate was most observant of the stay-at-home orders at approximately that date.

Figures \ref{fig:modularity} and  \ref{fig:intra} also suggest there is utility in simulating infectious disease spread on device graph datasets. The datasets provide weekly snapshots of contacts (connections) at country wide scale. The ability to simulate infectious disease spread on week over week snapshots allows closer to 'real time' estimations on the efficacy of stay-at-home orders; both at large and in specific areas of the US. In addition, being able to observe the resurgence of cross community edges week over week has forward looking utility.  It may be used to predict areas susceptible to an outbreak or resurgence of cases if the influx of cross community edges are observed in geographic areas still experiencing a steady rate of new infections. This data can allow policy makers to make better informed decisions on when it is safe to resume certain public functions and business.

\begin{table}[t]
	\centering
	\begin{tabular}{|c|c|c|}
		\hline 
	     & Nodes & Edges \\
	    \hline
		$G_0$ &  1,169,103,380 & 6,189,562,782 \\  \hline
		$G_{0.2}$ & 1,057,968,932 & 3,456,105,267 \\  \hline
		$G_{0.4}$ & 873,110,979 &  1,917,085,849 \\  \hline
		$G_{0.6}$ & 753,677,283  & 1,331,218,411 \\  \hline
		$G_{0.8}$ & 696,068,576 & 1,048,758,994 \\  \hline
		$G_{1.0}$ &   600,338,457 &  841,785,094 \\ \hline
	\end{tabular}
	\vspace{.15cm}
	\caption{Edge count and node count of the graph before `shelter-in-place'.  Data collected over six weeks from January 20th, 2020 to March 1st, 2020.  $G_{\gamma}$ corresponds to the graph, restricted to edge weights above $\gamma$.  }
	\label{tab:statsbefore}
\end{table}

\begin{table}[t]
	\centering
	\begin{tabular}{|c|c|c|}
		\hline 
	    & Nodes & Edges \\
	    \hline
		$G_0$ & 1,196,074,232 & 6,034,219,867 \\  \hline
		$G_{0.2}$ & 1,078,714,915 & 3,707,466,124 \\  \hline
		$G_{0.4}$ & 889,876,127 &  2,187,580,778 \\  \hline
		$G_{0.6}$ &  771,171,764 & 1,576,917,496 \\  \hline
		$G_{0.8}$ & 710,457,825 & 1,261,395,734 \\  \hline
		$G_{1.0}$ &  626,406,997 &  1,030,235,341 \\ \hline
	\end{tabular}
	\vspace{.15cm}
	\caption{Edge count and node count of the graph after `shelter-in-place'.  Data collected over six weeks from March 30th, 2020 to May 10th, 2020.  $G_{\gamma}$ corresponds to the graph, restricted to edge weights above $\gamma$.}
	\label{tab:statsafter}
\end{table}

%\begin{figure}[htb]
%\includegraphics[clip,scale=0.39]{figures/component_dist.pdf}
%\vspace{-.37cm}
%\caption{Before and after `shelter in place': connected component cumulative distribution.  The plots  show the fraction of notes in a connected component of size $x$ or smaller. }
%\label{fig:cc_dist}
%\end{figure}

\begin{figure}[htb]
\centerline{\includegraphics[clip,scale=0.55]{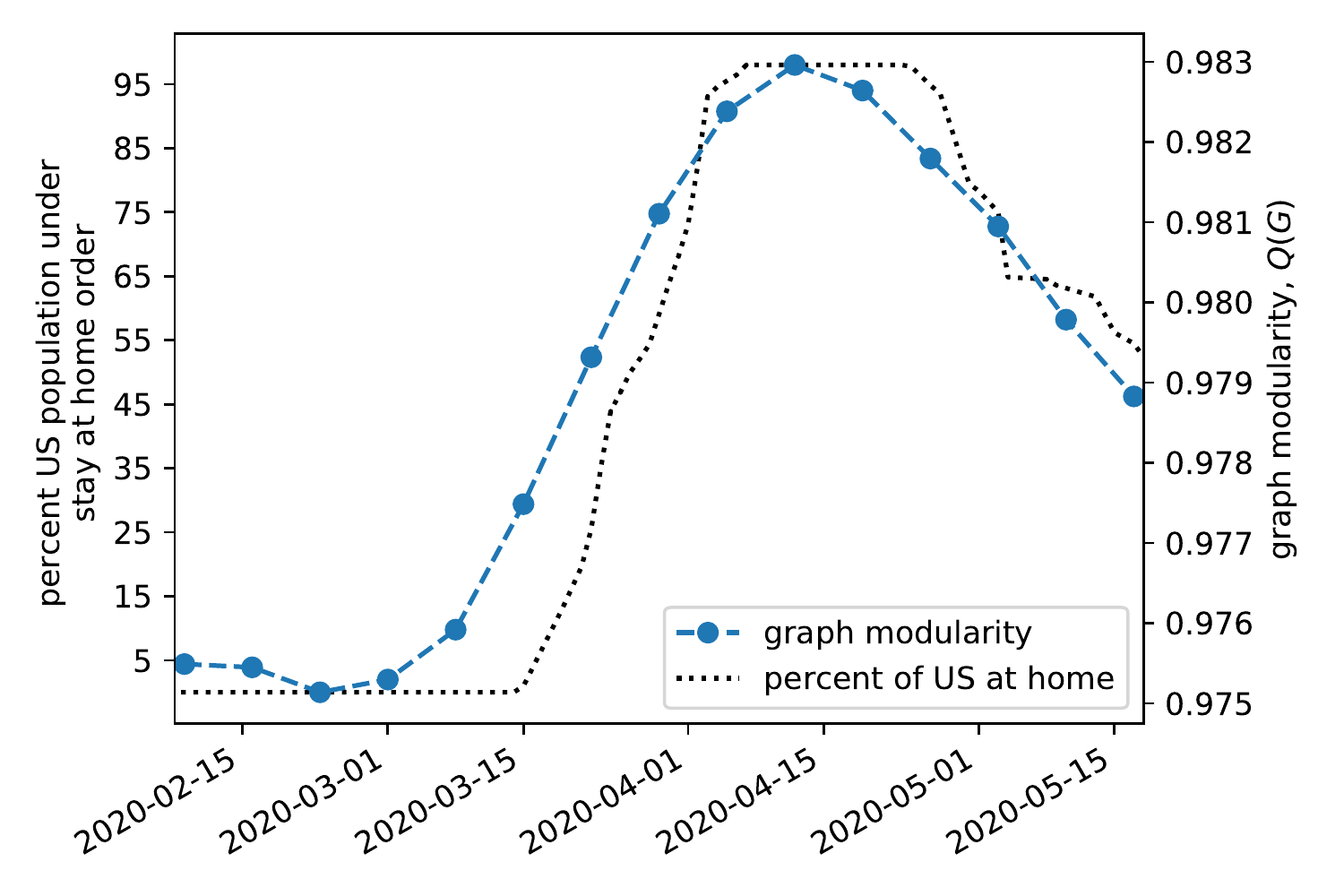}}
\vspace{-.37cm}
\caption{  Graph modularity (right axis), a measure of how well a graph matches community structure.  Percentage of US population under state-issued `stay-at-home' order (left axis, adapted from \cite{nytimes_stay_home}).  The graph is constructed with six weeks of data (three weeks before and three after the listed date).  }
\label{fig:modularity}
\end{figure}

\begin{figure}[htb]
\centerline{\includegraphics[clip,scale=0.55]{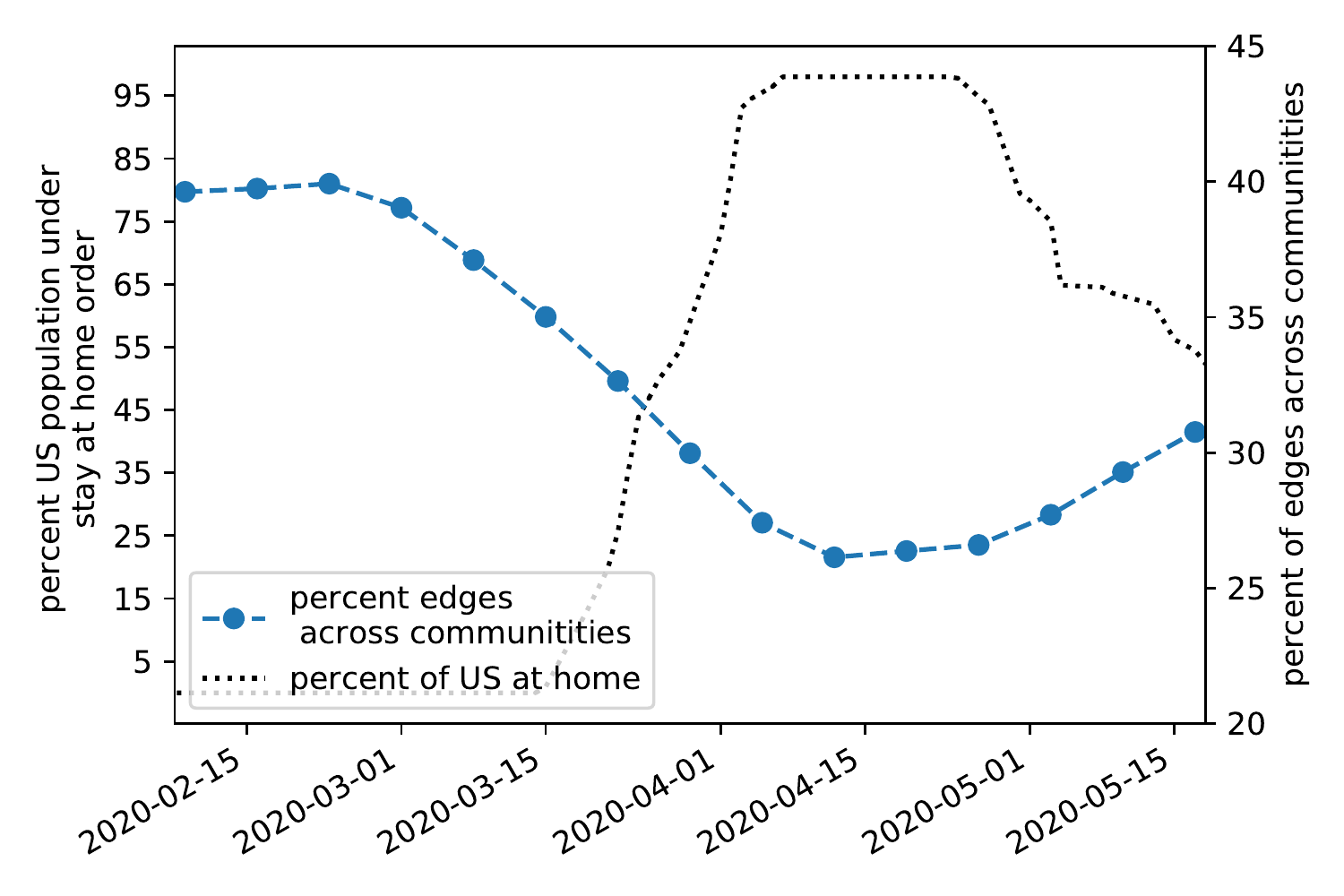}}
\vspace{-.37cm}
\caption{Percent of total edges that are inter-community (across community) edges (right axis). Percentage of US population under state-issued `stay-at-home' ordered (left axis, adapted from \cite{nytimes_stay_home}).  }
\label{fig:intra}
\end{figure}

Results of community detection were validated using a ground truthdataset provided by Comscore.  The datasets involves recruitment of \emph{panelists} for installation of customized wireless routers.  The wireless router captures an anonymized version of the media access control (MAC) address, and the cookies ID/advertising IDs of with the devices that connects to the router.

The cookie IDs and advertising IDs recorded by a custom wireless router define a ground truth community (which correspond to residential households).  The dataset includes more than 5,000 ground truth communities.  Only cookie IDs/advertising IDs that correspond to MAC addresses observed for more than 2 days are included, eliminating `guest' devices that do not have a long term association with the small community. The time frame of data collection used to define the ground truth communities is the same 6 weeks as the time frame used for graph data collection.

\emph{Precision} and \emph{recall} of communities produced by Louvain Modularity were calculated. In particular,  let $\mathcal{C}$ be the set of communities produced by the community detection algorithm. For a ground truth community $C'$ ({\em i.e.,}  a physical household), the \emph{best} corresponding graph community $C^*$  is determined by finding the community with the largest intersection of IDs:
$C^* := \arg \max_{C \in \mathcal{C}}   |C'  \cap C|$.
The precision and recall of $C^*$ are given by 
\begin{eqnarray*}
\mbox{precision} = \frac{ |C' \cap C^* |  }{|C^*|} \hspace{0.8cm} \mbox{recall} = \frac{ |C' \cap C^* |  }{|C' |} .
\end{eqnarray*}
The the precision and recall are plotted for Louvain Modularity applied to $G_{0.8}$ in Figure \ref{fig:pr}.  Notice that community structure is best aligned with a ground truth partitioning of the nodes between April 12th and April 19th, 2020. 

Figures \ref{fig:graph_before} and \ref{fig:graph_after} visualize a small portion of the graph, and represent an illustrative and hand-curated example of communities before and after shelter-in-place.  Notice that many connections across communities are not present after stay-at-home orders.  The example figures also highlight the utility of the graph with respect to contact tracing.

%Talk more about privacy concerns...

\begin{figure}[htb]
\centerline{\includegraphics[clip,scale=0.55]{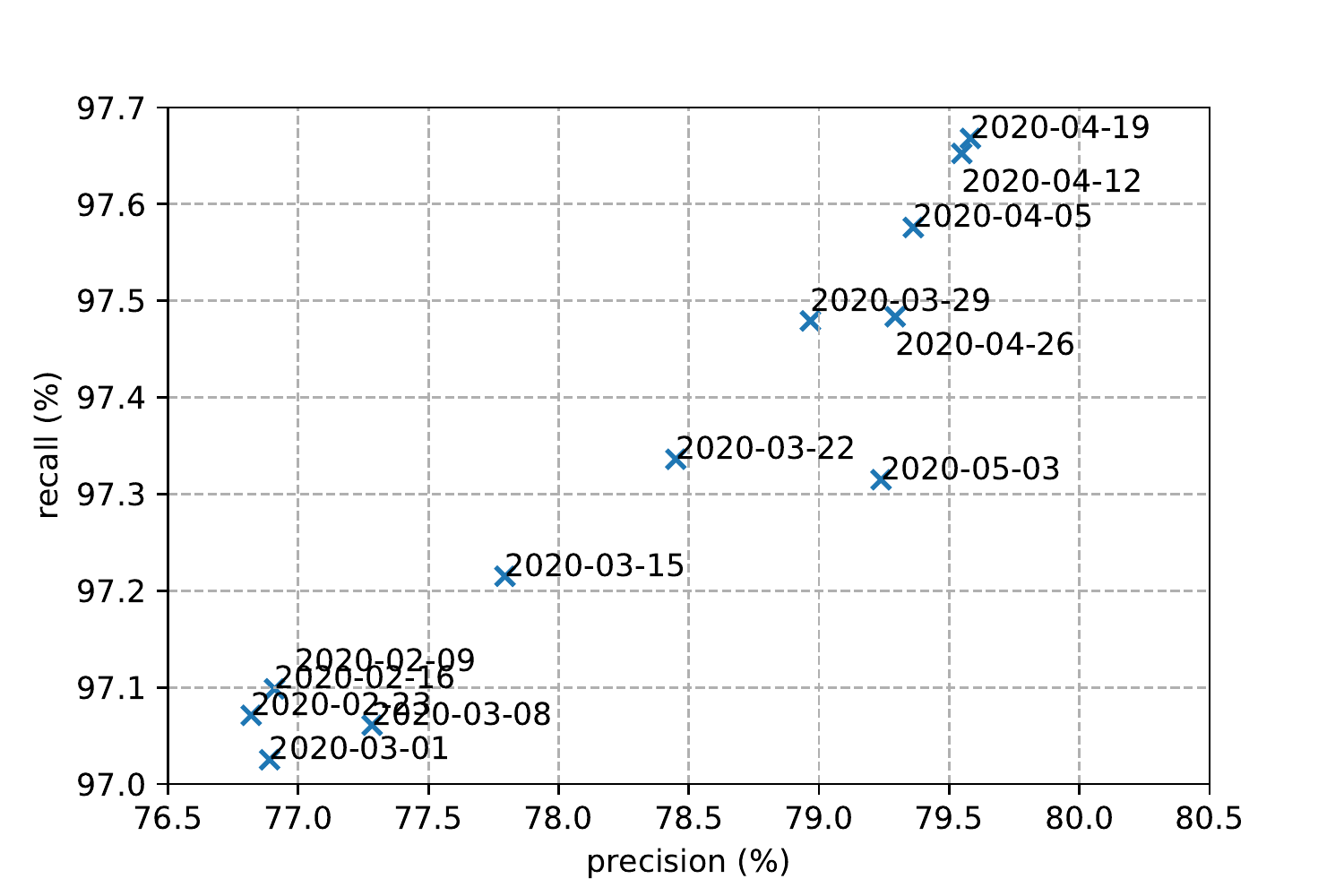}}
\vspace{-.37cm}
\caption{Precision and recall of communities derived from a ground truth dataset on more than 5000 communities.  }
\vspace{-.5cm}
\label{fig:pr}
\end{figure}

\begin{figure}[htb]
\centerline{\includegraphics[clip,scale=0.37]{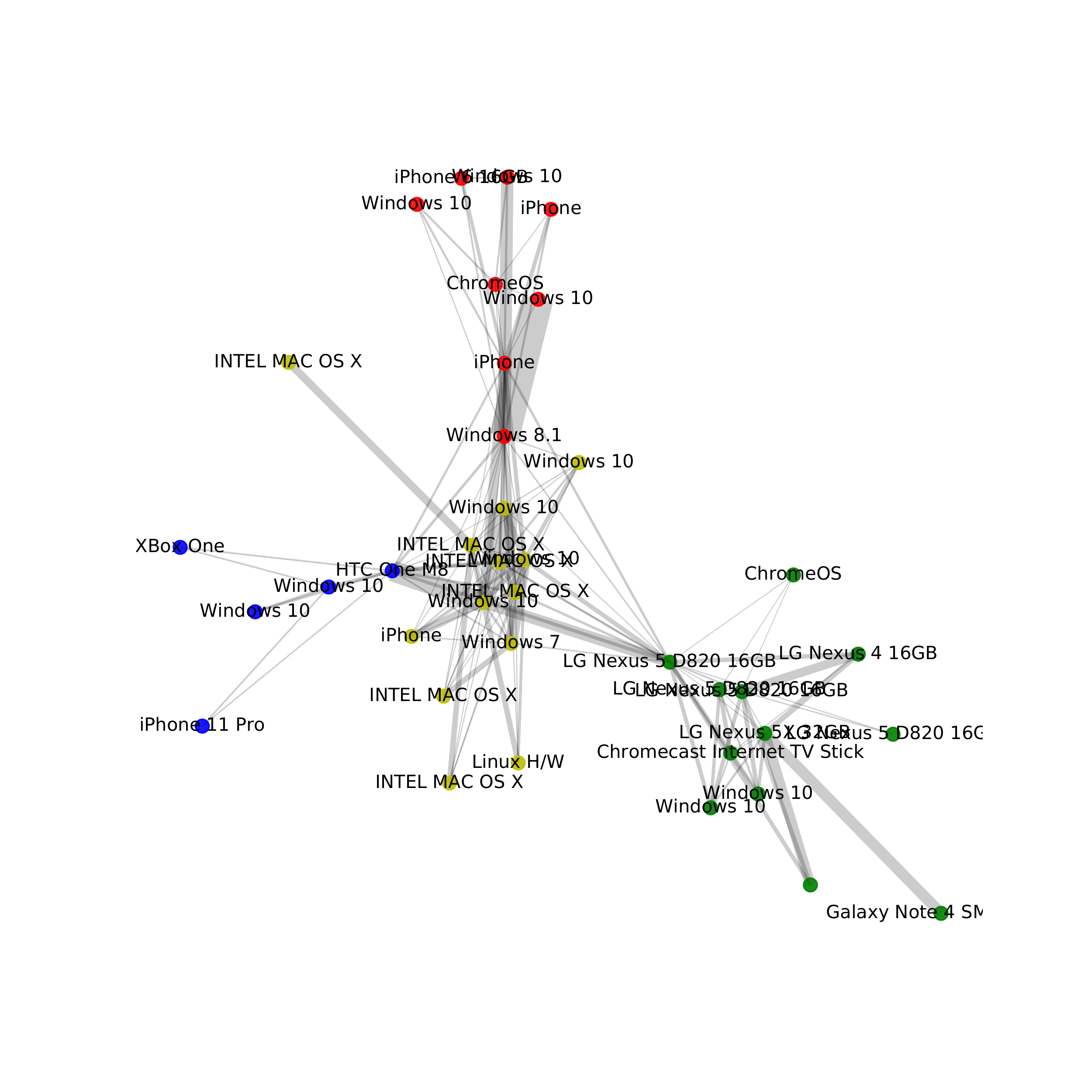}}
\vspace{-1.5cm}
\caption{Illustrative example of a small portion of the graph \emph{prior} to stay-at-home orders.  The color of node indicates the assigned community. Note the large number of connections between the device labeled `LG Nexus 5 D820 16GB' and devices outside its community.  }
\label{fig:graph_before}
\end{figure}

\begin{figure}[htb]
\vspace{-.2cm}
\centerline{\includegraphics[clip,scale=0.37]{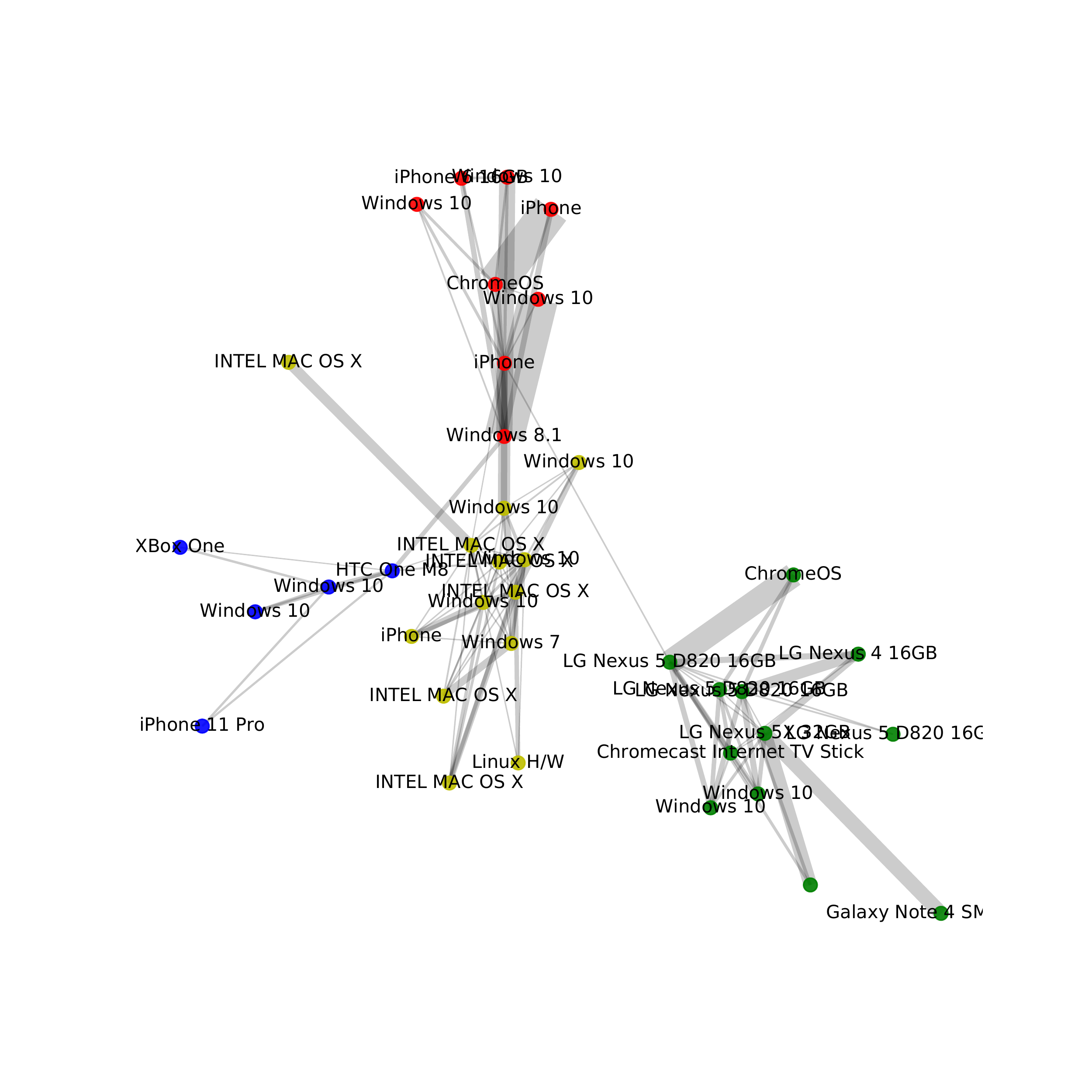}}
\vspace{-1.5cm}
\caption{Illustrative example of a small portion of the graph \emph{after} stay-at-home orders.  The color of node indicates the assigned community.  Notice that many of the edges outside the communities are no longer present.  }
\label{fig:graph_after}
\end{figure}

\section{Related Work}\label{sec:related_work} The study of networks (i.e, graphs) in epidemiology has a rich history dating back to at least the 1980's \cite{klovdahl1985social}.  There are a number of review articles on the topic of graphs and infectious disease, and while we aim to extract the relevant themes herein, we refer the reader to \cite{danon2011networks} for a more thorough discussion.

A number of studies have looked at how (offline) social networks predict, model and relate to the spread of infectious disease \cite{meyers2007contact, klovdahl1994social, klovdahl1985social}.  This includes studies of \emph{realized encounter networks}, in which a graph is constructed from documented physical encounters, and the spread of a transmitted disease is studied on the resulting graph.  A major limitation of such approaches is the construction of the graph itself, which requires participation from subjects and recollection of encounters.  Such studies have in general been limited to graphs with fewer than a thousand nodes.  One difference between our discussions and much work in epidemiology is that we do not have information about nodes are infected with a disease.  We acknowledge significant challenge in relating that information to the graph at scale.  Relating offline information (i.e, a name/address) to an advertising ID or cookie is often termed `digital onboarding', and a number of companies offer such datasets \cite{Liveramp}. 

In addition to the broader study of infectious disease on networks with social and encounter networks, there is work that studies the effects of graph modularity on the spread of infectious disease \cite{griffin2012community, sah2017unraveling, nunn2015infectious}.  While conventional wisdom would suggest that modular organization of a network would prevent the spread of infectious disease, there is work on animal social networks that in certain settings supposes otherwise \cite{sah2017unraveling}.  

There has been a surge in work related to digital contact tracing (see, for example \cite{ferretti2020quantifying} and related articles).  In general these techniques require a significant portion of the population install a mobile app for data collection, which is a significant challenge.  The technology involved is both based on global positioning (GPS), Bluetooth Low Energy \cite{gorji2020stecc, farrahi2014epidemic}.

%The work of \cite{} suggests using commnications 
%Data protection \cite{abeler2020covid}
%provide a 
%s To the best of our knowledge, there is no work relating device graphs to empidmiological graphs 
%\cite{khim2018theory, anderson1992infectious, christakis2007spread, dong2012graph, peterson2011contact}

Related work in device graphing is broadly divided into academic work, including the prequels to this paper \cite{Malloy17, Malloy18, Malloy19}, and commercial literature.  \cite{Malloy17, malloy2016ad} provides a introduction to device graphing using IP co-location, and is the first academic paper to study the datasets.   \cite{Malloy18} assumes that IP colocation associations have been made, and focuses on refining the graph to create custom \emph{user} and \emph{device} level groupings using an extension of Naive Bayes termed Naive Bayes Similarity Scoring.  Lastly, \cite{Malloy19} proposes an approach to extend the scale of the graph towards 10 billion nodes.  Beyond these publications, little academic literature exists with the taxonomy `device graph'.  Some literature on `cross device tracking'  \cite{solomos2018cross} aims to provide a method to detect when cross device tracking (such as device graphing) occurs.  A review of privacy in cross device tracking is found in \cite{zimmeck2017privacy}.

Commercial and industry publications with the `device graph' taxonomy are more prevalent, as device graph offerings are ubiquitous and widely available {\em e.g.,}~\cite{Tapad, Lotame, Liveramp} in the adverting ecosystem.  This includes a number of patents on systems for generation of device graphs based on colocation \cite{malloy2018systems, malloy2019systems, funkhouser2020reporting}.  The academic work can be viewed as a formalization of many of the ideas encapsulated in the commercial device graph patents and literature.   Much of the work that studies identification of users on the internet absent login information can be broadly categorized as \emph{fingerprinting} \cite{Eckersley10,Mowrey11}.  The goal of fingerprinting, as the name suggests, is to identify a user/browser in a persistent manner, absent login information and without cookies or advertising IDs.

%We note that while fingerprinting literature is broad, and 

%The Evercookie, while not a traditional browser cookie, stores data in browser accessible locations that cannot be easily cleared by a client~\cite{Kamkar10}.   Companies (including Adobe and Microsoft ~\cite{AdobeLSO,MSFT_MUID}) have built similar functionality, which has since been discontinued.    Other device fingerprinting techniques use features that are easily accessible at the browser~\cite{Eckersley10,Mowrey11} and device level (which is often termed cross-browser identification) ~\cite{boda2011user, cao2017browser}.  \emph{Canvas fingerprinting}, a technique in which the browser uses the HTML canvas to draw an object and then hashes the result to create a unique ID \cite{mowery2012pixel, fifield2015fingerprinting} has seen significant attention .  Cross-browser identification, which is limited to the the associated a user across browsers on the same machine, relies on either IP address-based \cite{boda2011user} or hardware and OS features \cite{cao2017browser}, and is known to have varying levels of accuracy~\cite{Nikiforakis13}.  

\section{Summary}\label{sec:summary} This paper proposes the utility of internet device graphs in epidemiology.  Device graphs are characterized by immense scale and ubiquitous use in digital advertising.   More precisely, this paper posits the utility of device graphs for four applications \emph{i)} contact tracing, \emph{ii)} prediction of `hot spots',  \emph{iii)} simulation of infectious disease spread, and \emph{iv)} delivery of advertisement based warnings to potentially exposed individuals.   The utility follows two natural observations \emph{i)} in epidemiology, the spread of infectious disease through a population can be modeled using a network or graph, and \emph{ii)} in digital advertising \emph{internet device graphs} are graph data sets that organize digital identifiers produced by mobile phones, PCs, TVs, and tablets as they access media on the internet.  

This paper also studies the effects of the shelter-in-place orders on a large scale internet device graph constructed using IP co-location with more than a billion nodes.   As individuals stay at home, fewer graph edges are created with co-workers and other individuals outside the home.   We study the impact on the community structure in the graph, and show that the graph modularity (a measure of how well a graph conforms to community structure) increases as the stay at home order went into place.  The modularity of the graph is closely related to the number of edges that cross communities, in relation to the total edges.  The number of cross community edges decreases as the stay at home order went into place.   Both metrics indicate that the graph is most modular (best adheres to community structure) the week beginning April 12th.  This implies that the US, in aggregate, best followed the stay at home orders that same week. 

While device graphs offer utility for the study of infectious disease, they may also supplement the more material and a immediate needs of public health departments that are actively fighting the worldwide COVID-19 pandemic.   The digital advertising ecosystem writ large has successfully built tools, infrastructure, and datasets to track users as they access services on the internet.  The question remains: can the processes and datasets amassed by the digital advertising industry help in the fight against COVID-19, and do so in a privacy safe manner?

\bibliographystyle{ACM-Reference-Format}
\balance
\bibliography{paper}

\end{document}